\documentclass[a4paper,fleqn,usenatbib]{mnras}
\usepackage[T1]{fontenc}
\usepackage{ae,aecompl}
\usepackage{graphicx}
\usepackage{amsmath}
\usepackage{amssymb}
\usepackage{txfonts}
\usepackage{bbold}
\usepackage{xcolor}

\usepackage{float}
\usepackage{adjustbox}

\graphicspath{ {images/} }

\newcommand{\ltsima}{$\; \buildrel < \over \sim \;$}
\newcommand{\lsim}{\lower.5ex\hbox{\ltsima}}
\newcommand{\gtsima}{$\; \buildrel > \over \sim \;$}
\newcommand{\gsim}{\lower.5ex\hbox{\gtsima}}

\newcommand{\dd}{\mathrm{d}}

\onecolumn

\title[surface brightness fluctuations]{Etherington duality breaking: gravitational lensing in non-metric spacetimes versus intrinsic alignments}
\author[E.S. Giesel, B. Ghosh, B.M. Sch{\"a}fer]
{Eileen Sophie Giesel$^1$\thanks{e-mail: e.giesel@stud.uni-heidelberg.de}, Basundhara Ghosh$^2$\thanks{e-mail: basundharag@iisc.ac.in}, Bj{\"o}rn Malte Sch{\"a}fer$^1$\thanks{e-mail: bjoern.malte.schaefer@uni-heidelberg.de}\\
$^1$Zentrum f{\"u}r Astronomie der Universit{\"a}t Heidelberg, Astronomisches Rechen-Institut, Philosophenweg 12, 69120 Heidelberg, Germany\\
$^2$Department of Physics, Indian Institute of Science, C. V. Raman Road, Bangalore 560012, India
}

\begin{document}
\pagerange{\pageref{firstpage}--\pageref{lastpage}}
\pubyear{2021}
\maketitle
\label{firstpage}

\begin{abstract}
The Etherington distance duality relation is well-established for metric theories of gravity, and confirms the duality between the luminosity distance and the angular diameter distance through the conservation of surface brightness. A violation of the Etherington distance duality due to lensing in a non-metric spacetime would lead to fluctuations in surface brightness of galaxies. Likewise, fluctuations of the surface brightness can arise in classical astrophysics as a consequence of intrinsic tidal interaction of galaxies with their environment. Therefore, we study these in two cases in detail: Firstly, for intrinsic size fluctuations and the resulting changes in surface brightness, and secondly, for an area-metric spacetime as an example of a non-metric spacetime where the distance duality relation itself acquires modifications. The aim of this work is to quantify whether a surface brightness fluctuation effect due to area-metric gravity would be resolvable compared to the similar effect caused by intrinsic alignment. We thus compare the auto- and cross-correlations of the angular spectra in these two cases and show that the fluctuations in intrinsic brightness can potentially be measured with a cumulative signal-to-noise ratio $\Sigma(\ell) \geq 3$ in a Euclid-like survey. The measurement in area-metric spacetimes, however, depends on the specific parameter choices, which also determine the shape and amplitude of the spectra. While lensing surveys do have sensitivity to lensing-induced surface brightness fluctuations in area-metric spacetimes, the measurement does not seem to be possible for natural values of the Etherington-breaking parameters.
\end{abstract}

\begin{keywords}
gravitational lensing: weak -- dark energy -- large-scale structure of Universe -- gravitation -- cosmology: theory
\end{keywords}

\section{Introduction}\label{sect_intro}
The distance duality relation given by Etherington almost a hundred years ago, has established a unique characteristic of metric theories of gravity \cite[Chap. 3]{Schneider1992}, and connects two important cosmological distances by a simple relation as a function of the redshift $z$:
\begin{equation}
    D_L(z)=D_A(z)(1+z)^2,
\label{e:dist_dual}
\end{equation}
$D_L(z)$ being the luminosity distance relevant for standard candles, and $D_A(z)$ being the angular diameter distance for standard rulers. The relation in Eq.~\ref{e:dist_dual} tells us that for any metric theory of gravity where photons follow null-geodesics and the photon number is conserved, any of the above two cosmological distances can be used to make measurements since they are dual to each other. In literature, there are many instances where the violation of this distance-duality relation has been studied, and has been used to suggest the existence of new physics or deviations from metric theories of gravity  \citep[see for instance][]{Bassett_2004, schuller2017etheringtons, more2016modifications}. A violation of the Etherington distance duality itself leads to a change in the surface brightness conservation law in gravitational lensing. 

A particular class of non-metric theories of gravity arise in the context of gravitational closure, which itself is based on the idea that one can systematically construct a theory of gravity given a specific matter action and the corresponding causal structure of the spacetime as inputs \citep[see][for more details]{schneider2017gravitational,Closure2018,Giesel_2012,Witte2014Thesis,Duell2020thesis,Wolz2022thesis}, thus extending ideas of classical geometrodynamics \citep[see][and references therein]{geometro76,Kuchar74}. One can for instance show that in this framework the action for standard Maxwellian electrodynamics enables to actually construct the Einstein-Hilbert action for the gravitational interaction. But the spacetime structure does not necessarily have to be a metric, instead, one could provide a measure of area, which typically arises, in general linear electrodynamics in a pre-metric setting \citep{Hehl_2003, Obdukov_2002, Rubilar_2002} via
\begin{equation}
S_{\text{matter}} \left[A;G\right) = 
-\frac{1}{8}\int\dd^4 x\: \omega_{G}\:  G^{abcd}  F_{ab} F_{cd}.
\end{equation}
Here, $G^{abcd}$ is the so-called area-metric or constitutive tensor, which is a $4$-rank tensor with the same symmetries as the Riemann tensor. It replaces the metric, which gives a measure of length for vectors by a general measure of areas of parallelograms spanned by pairs of vectors. Also, the volume element of such a space-time is defined with a scalar density factor $\omega_{G}=1/24 \left(\epsilon_{abcd} G^{abcd}\right)^{-1}$, instead of the square root of the determinant of a metric acting as a covolume.

In an axiomatic approach one can show that the corresponding electrodynamical theory allows in principle for vacuum birefringence \citep[see][for instance]{Punzi:2007di,Rivera_2011}. The dispersion relation for light propagation is hereby defined via the $4$-rank Fresnel tensor $P^{abcd}$ or principal polynomial
$P^{abcd}k_a k_b k_c k_d = -\frac{1}{24}\omega_{G}^2\epsilon_{u v p q}\epsilon_{r s t u} G^{u v r(a} G^{b|ps|c} G^{d)qtu} k_a k_b k_c k_d$,
which is generally non-metric in this case \citep[see][for more details and references therein]{Raetzel_2011}.

Even though the idea of an area-metric might seem exotic, such concepts already appear in the context of classical electrodynamics in matter. Here, the constitutive tensor defines a linear relationship between the electromagnetic field in vacuum and the induced fields due to interaction with matter. While finding a closed solution for a fully area-metric theory of gravity is still subject of ongoing research \citep{Closure2018,Wolz2022thesis}, one can already find solutions for highly symmetric spacetimes \citep{Duell2020thesis} and in the weak field limit \citep{schneider2017gravitational,Wolz2022thesis,schuller2017etheringtons, Alex_Proceedings}. The violation of the distance-duality relation in area-metric theories of gravity can be understood by considering the Gotay-Marsden source tensor (for details refer to \citet{Gotay1992StressEnergyMomentumTA}) as a generalisation of the energy-momentum tensor in arbitrary geometry. Most importantly, as discussed in more detail in \citet{schuller2017etheringtons, Fischer2017Thesis}, one may recover a notion of covariant energy-momentum conservation for area-metric spacetimes as well, however with respect to a different volume element $\omega_G$ as according scalar density factor for an area-metric geometry, instead of $\sqrt{-\text{det}(g)}$ as in the case of metric geometry. For a weakly birefringent space-time, discussed by \citet{schuller2017etheringtons}, with effectively metric geodesics for light propagation, which occurs since the according principal polynomial is in this case metric to first order, one finds a photon excess factor when integrating the photon current between source and observer. This happens because then the covariant derivative with respect to the effectively metric principal polynomial of the photon current vector $N^a$ is non-vanishing with $\nabla_a N^a \neq 0$, while the actual covariant photon current density conservation is now given by $\partial_a\left(\omega_{G} N^a\right) = 0$. Thus, because of this effective photon excess a violation of the Etherington distance duality would occur in this case \citep[see][for more details for instance]{schuller2017etheringtons}, which leads to a violation of the surface brightness conservation law, and thus surface brightness fluctuations, in gravitational lensing. 

Thinking of gravitational weak lensing effects in the shapes and sizes of galaxies as light sources one could invoke a surface brightness fluctuation, which is otherwise non-existent in metric theories of gravity and which would provide hints at non-metric structures, in a purely conventional gravitational theory, and even in Newtonian gravity through intrinsic alignments \citep[see][]{joachimi_galaxy_2015, kirk_galaxy_2015,kiessling_galaxy_2015, troxel_intrinsic_2015}, i.e. through direct tidal gravitational interaction of a galaxy with its environment. The trace of the tidal field has the capability of physically compressing a galaxy \citep{ghosh2020intrinsic}, leading to fluctuations in angular size. With the light emission being unaffected, the surface brightness would then increase as a consequence of the tidal interaction, such that even in a Newtonian theory of gravity one would expect surface brightness fluctuations through this particular astrophysical mechanism.

The aim of this work is hence to derive the statistical spectra for both effects and to estimate their respective observabilities on different scales, to evaluate which effect would be dominant.

The paper is structured as follows: In section \ref{sect_fluctuations_intrinsic} we shortly summarise the linear alignment model for elliptical galaxies introduced by \citet{ghosh2020intrinsic}, show how this model leads to intrinsic surface brightness fluctuations and find the corresponding spectra. Then, we discuss in section \ref{sect_fluctuations_etherington} how the Etherington distance duality violation derived by \citet{schuller2017etheringtons} leads to a surface brightness fluctuation spectrum in an area-metric setting. Finally, we evaluate these spectra numerically in section \ref{sect_numerical} and summarise our results and discuss limits to our approach in \ref{sect_summary}. We assume a $w$CDM-cosmology with standard cosmological parameter values as $w$ close to $-1$ ($w=-0.9$), $\Omega_m = 0.3$, $\sigma_8 =  0.8$, $h = 0.7$ and $n_s = 0.96$, and use Einstein sum convention throughout the paper.

\section{Fluctuations of intrinsic surface brightnesses}
\label{sect_fluctuations_intrinsic}
Intrinsic alignments affect the intrinsic surface brightness $I$ of a source, here defined as the logarithmic surface brightness $S$ of e.g. a galaxy \citep{Freudenburg:2019dyh} as 
\begin{equation}\label{eq:surfacebrightnessDefinition}
S = \log_{10}\left(\frac{F}{A}\right) = \log_{10}\left(I\right)
\end{equation}
where $F$ is the flux density and $A$ is the intrinsic area of the galaxy in terms of a solid angle. While the image surface brightness is generally known to be - up to a redshift correction factor - equivalent to the intrinsic surface brightness in gravitational lensing in general relativity \citep[Chap. 3]{Schneider1992}, the influence of local tidal fields $\Delta \Phi$ will lead to an effective change in the physical size of a galaxy and consequently of its cross section area $A$. In contrast, the number of stars $N_{\text{stars}}$ and, due to a simple proportionality $N_{\text{stars}} \varpropto F$, the flux density of the galaxy is assumed to stay the same under the influence of tidal fields. Combining the flux that remains independent of $\Delta\Phi$ with the size which depends on $\Delta\Phi$ leads directly to a change in surface brightness which reflects the trace of the tidal fields: This result is in contrast to the conserved surface brightness in gravitational lensing on a metric spacetime.

We derive the effect of surface brightness fluctuations due to intrinsic alignment for standard Newtonian gravity, and assume that effects of possible area-metric corrections to gravity, as discussed in the next section, will be suppressed on local scales, and only become relevant on the Hubble-scales. Furthermore, any corrections due to area-metric gravity will effectively be contained in the empirical alignment parameter $D_{\text{IA}}$ and are thus hardly separable, especially since the numerical precision of the alignment parameter is determined up to a factor of $10$ \citep{zjupa2020intrinsic,Tugendhat_2018}. Also, we want to estimate the magnitude of this purely classical effect to later compare it to a possible effect of an exotic spacetime structure: This will turn out to be a curious case, as intrinsic alignment effects will dominate over gravitational lensing. As a last remark it is worth noting that $\Delta\Phi$ as the trace of the tidal gravitational fields could be responsible for an increased or decreased accretion onto the galaxy (depending on the sign of $\Delta\Phi$), which could affect star formation and therefore indirectly the flux $F$: We neglect this particular complication.

With a linear alignment model \citep[see][who first introduced this idea]{hirata_galaxy-galaxy_2004,hirata_intrinsic_2007,hirata_intrinsic_2010} for elliptical galaxies discussed by \citet{ghosh2020intrinsic} we can analytically derive the changes in size and thus cross section area: The linear alignment model hereby assumes that, according to \citet{piras_mass_2018}, one can use the Jeans-equation to describe how variations in the stellar density $\rho(r)$ of a galaxy as virialised system with a constant velocity dispersion $\sigma^2$ can be related to variations in the gravitational field $\Phi$ via
\begin{equation}
\sigma^2 \partial_r \ln\left( \rho(r) \right) = - \partial_r \Phi.
\end{equation}
The according solution for the stellar density is given by
\begin{equation}\label{eq:JeansDensity}
\rho(r) = \bar{\rho} \exp\left( - \frac{\Phi(r)}{\sigma^2}  \right) = \frac{N_{\text{stars}}}{A} \exp\left( - \frac{\Phi(r)}{\sigma^2}  \right),
\end{equation}
which is considered to be proportional to the according surface brightness distribution $I(r)$. 
Here $\bar{\rho}=N_{\text{stars}}/A$ is the normalization of the stellar density in units of stars per squared arc second. Since $F \varpropto N_{\text{stars}}$ this factor would thus be proportional to the normalization $\int \mathrm{d}^2 r \: I\left(r \right)=\bar{I}=F/A$ of the surface brightness distribution $I(r)$. 
The total number of stars - proportional to the flux - and thus the size $A$ is given by
\begin{equation}
\begin{split}
&N_{\text{stars}} =  
\int \mathrm{d}^2 r\: \rho\left(r \right) = \int \mathrm{d}^2 r\: \frac{N_{\text{stars}}}{A} \exp\left( - \frac{\Phi(r)}{\sigma^2}  \right) \,\,\Rightarrow \,\, & 
A = \int \mathrm{d}^2 r\:  \exp\left( - \frac{\Phi(r)}{\sigma^2}  \right) = 
\int_{0}^{2 \pi} \mathrm{d} \phi\: \int_{0}^{\infty} \mathrm{d}r r\:  \exp\left( - \frac{\Phi(r)}{\sigma^2}  \right).
\end{split}
\end{equation}
Now perturbations of the potential due to the influence of local tidal fields according to 
\begin{equation}\label{eq:perturbedPotential}
\Phi(r) \, \rightarrow \,  \Phi(r) + \frac{1}{2} \partial_{a} \partial_{b} \left.\Phi\right\vert_{r=0} r^a r^b,
\end{equation}
\citep{ghosh2020intrinsic} lead to modifications in the stellar density according to
\begin{equation}\label{eq:NewDensity}
\rho^\prime(r) =  
\bar{\rho}^\prime \exp\left(-\frac{\Phi(r)}{\sigma^2}  \right) \left(1 - \frac{1}{2 \sigma^2} \partial_{a} \partial_{b} \Phi\: r^a r^b \right) = \frac{N_{\text{stars}}}{A^\prime} \exp\left(-\frac{\Phi(r)}{\sigma^2}  \right) \left(1- \frac{1}{2 \sigma^2} \partial_{a} \partial_{b} \Phi\: r^a r^b \right) .
\end{equation}
Here the tidal field $\partial_{a} \partial_{b} \Phi$ is evaluated at the center of the galaxy set at $r=0$, and the indices $a$ and $b$ run from $0$ to $1$ with polar coordinates $r_0 = r \cos \phi $ and $r_1 = r \sin \phi$.
While the total number of stars does not change as discussed before, the intrinsic cross section galaxy area $A^\prime$ in the normalization factor $\bar{\rho}^\prime=N_{\text{stars}}/A^\prime$ needs to change according to 
\begin{equation}
\begin{split}
A^\prime = \int \mathrm{d}^2 r\:  \exp\left( - \frac{\Phi(r)}{\sigma^2}  \right) \left(1- \frac{1}{2 \sigma^2} \partial_{a} \partial_{b} \Phi r^a r^b \right) = 
\int_{0}^{2 \pi} \mathrm{d} \phi\: \int_{0}^{\infty} \mathrm{d}r\: r  \exp\left( - \frac{\Phi(r)}{\sigma^2}  \right) \left(1- \frac{1}{2 \sigma^2} \partial_{a} \partial_{b} \Phi r^a r^b \right) .
\end{split}
\end{equation}
Thus, there is a relative variation in the stellar density $\bar{\rho}^\prime-\bar{\rho}/\bar{\rho} \equiv \delta \bar{\rho}/\bar{\rho}$, and hence size or rather cross section area given by $\left(A-A^\prime\right)/A^\prime \equiv - \delta A/A^\prime$. Due to (\ref{eq:surfacebrightnessDefinition}) this leads to a relative variation in the surface brightness distribution since
\begin{equation}
\label{eq:surfacebrightnessvariation}
\frac{\delta \bar{\rho}}{\bar{\rho}}= \frac{N_{\text{star}}/A^\prime-N_{\text{star}}/A}{N_{\text{star}}/A} = \frac{-\delta A}{A^\prime}= \frac{F/A^\prime-F/A}{F/A}  = 
\frac{\bar{I}^\prime-\bar{I}}{\bar{I}} = \frac{\delta \bar{I} }{\bar{I}} = \frac{1}{2 \sigma^2} \frac{\int_{0}^{2 \pi} \mathrm{d} \phi\: \int_{0}^{\infty} \mathrm{d}r\: r \exp\left( - \Phi(r)/\sigma^2 \right)  \partial_{a} \partial_{b} \Phi r^a r^b }{\int_{0}^{2 \pi} \mathrm{d} \phi\: \int_{0}^{\infty} \mathrm{d}r\: r  \exp\left( - \Phi(r)/\sigma^2  \right)} .
\end{equation} 
Explicitly inserting a Sérsic model for the density profile (see \citet{sersic_influence_1963, graham_concise_2005, de_vaucouleurs_recherches_1948} for more details) 
\begin{equation}\label{eq:sersic}
\rho(r)\varpropto \exp\left(-\frac{\Phi(r)}{\sigma^2}\right) \hat{=} \exp\left(-b(n) \left[\left(\frac{r}{r_{\text{scale}}}\right)^{1/n} -1   \right] \right)\, \quad \text{with $b(n) \approx 2n -\frac{1}{3}$ and $n$ the Sérsic index},
\end{equation}
similar to previous work on linear intrinsic alignment \citep{ghosh2020intrinsic,Giesel_2021}, and using the substitutions $x=b\left[\left(r / r_{\text{scale}}\right)^{n^{-1}}-1 \right]$ and $y=x+b$ one can rewrite the integrals in expression (\ref{eq:surfacebrightnessvariation}) in terms of Gamma functions $\Gamma(n)$ of the Sérsic index $n$. 

We thus find
\begin{equation}
\int_{0}^{2 \pi} \mathrm{d} \phi\: \int_{0}^{\infty} \mathrm{d}r\: r  \exp\left( - \Phi(r)/\sigma^2  \right) = \frac{2\pi n}{b^{2n}} \exp(b) r_{\text{scale}}^2 \Gamma(2n),
\end{equation}
as well as
\begin{equation}
\int_{0}^{2 \pi} \mathrm{d} \phi\: \int_{0}^{\infty} \mathrm{d}r\: r \exp\left( - \Phi(r)/\sigma^2 \right)  \frac{\partial_{a} \partial_{b} \Phi}{2 \sigma^2} r^a r^b = 
\frac{2\pi n}{4 b^{4n} \sigma^2} \exp(b) r_{\text{scale}}^4 \Gamma(4n)  \Delta \Phi,
\end{equation}
such that the intrinsic surface brightness fluctuation is given by
\begin{equation}
\label{eq:IntrinsicIFin}
\frac{\delta \bar{I}}{\bar{I}}= \frac{1}{4}r_{\text{scale}}^2 b^{-2n}\frac{ c^2}{\sigma^2} \frac{\Gamma(4n)}{\Gamma(2n)} \frac{\Delta \Phi}{c^2} = 4 \frac{ \Gamma(4n)^2}{\Gamma(2n) \Gamma(6n)} D_{IA} \frac{\Delta \Phi }{c^2}=  2 S_{\text{Sérsic}}(n) \Delta s . 
\end{equation}
Here, the alignment parameter $D_{IA}$ estimated by \citet{ghosh2020intrinsic} as
\begin{equation}
 D_{IA} \varpropto \frac{1}{2} \frac{c^2}{\sigma^2} \frac{\int \mathrm{d}^2 r\: \rho(r) r^4}{\int \mathrm{d}^2 r \rho(r) r^2} = 
 \frac{1}{2} \frac{c^2}{\sigma^2} \frac{1}{8} r_{\text{scale}}^2 b^{-2n} \frac{\Gamma(6n)}{\Gamma(4n)},
\end{equation}
was inserted and the Gamma functions were summarised as $S_{\text{Sérsic}}(n) \equiv 4\Gamma(4n)^2/(\Gamma(2n) \Gamma(6n))$. The exact numerical value for the alignment parameter is not known, but from simulations one knows that it should be settled between $D \simeq 10^{-4}~(\text{Mpc/h})^2$ \citep[see][for instance]{Tugendhat_2018,Hilbert_2017} to $D \simeq 10^{-6}~(\text{Mpc/h})^2$ \citep{zjupa2020intrinsic}. However, since the linear alignment model, and thus the according velocity dispersion $\sigma$, with $\sigma = 100~ \text{km}/\text{s}$ for systems like the Milky Way to scale the alignment parameter, is based on a virialised system, we need to further rescale the alignment parameter by a factor $r_{\text{vir}}^2/r_{\text{scale}}^2 = 10^4$ with Sérsic scale radius $r_{\text{scale}}=  2~\text{kpc}$ and virial radius $r_{\text{vir}}=  200~\text{kpc}$. Thus, we use a numerical value of $D_{\text{IA}} \simeq c^2/\sigma^2 10^{-2}~\text{Mpc/h}^2$ for our final alignment parameter modulo Sérsic pre-factors which do not have a strong effect on the magnitude of the parameter \citep[see][]{ghosh2020intrinsic}.

Now relation~(\ref{eq:IntrinsicIFin}) shows that the intrinsic surface brightness fluctuation is proportional to $\Delta \Phi/c^2$, similarly to the intrinsic size fluctuation $\Delta s \equiv \delta s/s $ derived by \citet{ghosh2020intrinsic} as $\Delta s= 1/2 D_{IA} \Delta \Phi /c^2$, which is not surprising since we directly argued at the beginning that the source of intrinsic surface brightness fluctuations are intrinsic size changes. We can read off from relations (\ref{eq:surfacebrightnessvariation}) and (\ref{eq:IntrinsicIFin}) that an over-dense region with $\Delta \Phi>0$ increases $\delta \bar{I}/{I}>0$ while the cross sectional area measure of the galaxy should decrease with $-\delta A/A^\prime<0$. This also holds for the intrinsic size $\Delta s$ as intrinsic volume measure defined in terms of fluctuations of the second moment of the surface brightness distribution $q_{ab} = \int \mathrm{d}^2 r\: I(r) r_a r_b$ derived by \citet{ghosh2020intrinsic}. Finally, the $II$-spectrum $C_{A B}^{\delta \bar{I}/\bar{I} \, \delta \bar{I}/\bar{I}}(\ell)$ for the auto-correlation of intrinsic surface brightness fluctuations in a tomographic analysis is given in terms of the intrinsic size spectrum by \citet{ghosh2020intrinsic} as
\begin{equation}\label{eq:IIspectrumE}
II: \,\, C_{A B}^{\delta \bar{I}/\bar{I} \, \delta \bar{I}/\bar{I}}(\ell) = 4 S_{\text{Sérsic}}(n)^2  C_{A B}^{\Delta s \, \Delta s}(\ell) = \ell^4 S_{\text{Sérsic}}(n)^2 D_{\text{IA}}^2 \int_0^{\chi_H} \frac{\mathrm{d} \chi}{\chi^2}\: W_{\varphi,A} W_{\varphi,B} P_{\Phi \Phi}(k=\ell/\chi),
\end{equation}
where Limber approximation \citep{limber} is applied. Also, the spatial derivatives are expressed in terms of the angular derivatives via $\partial_{r_a} \equiv \partial_{\theta_a} \chi^{-1}$ , with comoving distance $\chi$. With $\partial_{\theta_a} \rightarrow - \text{i} \ell_a$ in Fourier space and the assumption of statistical isotropy of the modes the Laplacian then transforms to $\Delta \rightarrow - \ell^2/\chi^2$, where $\chi^{-2}$ later goes into the definition of the weighting function (\ref{eq:weight}).
Here $P_{\Phi \Phi}(k=\ell/\chi) \varpropto k^{n_S-4} T(k)^2$ is the power spectrum for potential fluctuations in comoving coordinates with transfer function $T(k)$ extended to non-linear structure formation according to \citet{Smith2003} and $n_S\leq1$ the spectral index. The according weighting function is given by
\begin{equation}
\label{eq:weight}
W_{\varphi,A} \left(\chi\right) =  \frac{1}{\chi^2}p\left(z(\chi)\right) \Theta\left(\chi - \chi_A \right) \Theta\left(\chi_{A+1} - \chi\right)  \frac{\mathrm{d} z}{\mathrm{d}\chi^\prime} \frac{D_+(a)}{a},
\end{equation}
\citep{ghosh2020intrinsic} with the Hubble-law as $- \mathrm{d} \chi H(\chi) = c\mathrm{d} z $, the linear growth factor $D_+ (a)/a$ giving the time evolution of the potential, $\Theta\left(\chi - \chi_A \right)$ and $\Theta\left(\chi_{A+1} - \chi\right)$ as Heaviside functions to restrict the alignment effect to a local bin $A$, and finally $p\left(z(\chi)\right)$ as galaxy redshift distribution proposed by \citet{laureijs2011euclid}
\begin{equation}
\label{eq:eucliddistribution}
p(z)\varpropto \left( \frac{z}{z_0} \right)^2 \exp \left[- \left( \frac{z}{z_0} \right)^\beta \right] \quad \text{with $\beta = 3/2$ and $z_0 = 0.64$.}
\end{equation} 


\section{Fluctuations of surface brightnesses for a modified Etherington Distance Duality Relation}
\label{sect_fluctuations_etherington}
\citet{schuller2017etheringtons} consider a weakly birefringent spacetime, where the area-metric is given by a Minkowskian background geometry $G^{abcd} = \eta^{ac} \eta^{bd} - \eta^{ad} \eta^{bc} - \sqrt{- \text{det} \eta}\: \epsilon^{abcd} + H^{abcd}$ with purely area-metric perturbation $H^{abcd}$, which can be sourced by a point mass for instance. In this case the components of this perturbation include Yukawa-like correction terms to a Newtonian potential such that the effective potential can be expressed as \citep[see also][]{Alex_Proceedings,rieser2020thesis}
\begin{equation}\label{eq:effPotential}
\Phi_{\text{eff}}\left(\bmath{r}\right) = - \frac{G M}{c^2 \left\vert \bmath{r}-\bmath{r}_M  \right\vert}\left(1 +\delta \exp\left(-\eta  \left\vert \bmath{r}-\bmath{r}_M \right\vert \right) \right),
\end{equation}
where $\bmath{r}_M$ is the position of the point mass set between source and observer and $\left\vert \bmath{r}-\bmath{r}_M \right\vert$ a Euclidean distance \citep[for a detailed derivation of this result via gravitational closure or by different formalisms, see][]{Closure2018, schneider2017gravitational, Alex_Proceedings, Alex_2020}. Here, the constants $\eta$ as inverse range of the Yukawa interaction and $\delta$ as coupling to the Newtonian potential appear as constants in the area-metric refinement and their values will be discussed later in section \ref{sect_numerical}. Yukawa corrections to the Newtonian potential appear in many versions of $f(R)$ theories \citep[see][for instance]{CLIFTON20121,amendola_tsujikawa_2010}, however in area-metric geometries one additionally receives an effective photon excess fraction $\mu_{\text{vio}}$ when integrating the photon number from source to observer with respect to an effective metric volume element, while the volume element with respect to which the photon flux density is conserved is actually area-metric. For an area-metric spacetime with a weak curvature sourced by a point mass \citet{schuller2017etheringtons} find 
\begin{equation}
\mu_{\text{vio}}=\frac{3 \delta G M}{c^2}\left(\frac{\exp\left(-\eta \vert \bmath{r}_{\text{S}} -\bmath{r}_{\text{M}} \vert\right)}{\vert \bmath{r}_{{S}} -\bmath{r}_{{M}} \vert}-\frac{\exp\left(-\eta \vert \bmath{r}_{{O}} -\bmath{r}_{{M}} \vert\right)}{\vert \bmath{r}_{{O}} -\bmath{r}_{{M}} \vert}\right),
\label{eqn_magic} 
\end{equation}
for the photon excess fraction, with $\vert \bmath{r}_{{S}} -\bmath{r}_{{M}} \vert$ as the distance between point mass $M$ and source $S$ and $\vert \bmath{r}_{{O}} -\bmath{r}_{{M}} \vert$ respectively between the mass and the observer $O$. Thus, there is an effective flux enhancement $F^\prime = F \left(1+\mu_{\text{vio}}\right)$. Due to the definition of the bolometric luminosity as $L= 4 \pi D_L^2 F $, which should be constant since the actual luminosity of the source is not affected by the area-metric geometry, this leads to a rescaling of the luminosity distance as $D_L^\prime = D_L/ \sqrt{1+\mu_{\text{vio}}}$.
Thus the modified Etherington distance duality would in general be given by \citep{schuller2017etheringtons} 
\begin{equation}\label{eq:Etherington}
D_L(z) =  \frac{\left(1+z\right)^2 D_A(z)}{\sqrt{1+\mu_{\text{vio}}}}\approx \left(1+z\right)^2 D_A(z) \left(1+\frac{3 \delta G M}{2 c^2}\left(\frac{\exp\left(-\eta \vert \bmath{r}_{\text{S}} -\bmath{r}_{{M}} \vert\right)}{\vert \bmath{r}_{{S}} -\bmath{r}_{{M}} \vert}-\frac{\exp\left(-\eta\vert \bmath{r}_{{O}} -\bmath{r}_{{M}} \vert\right)}{\vert \bmath{r}_{{O}} -\bmath{r}_{{M}} \vert}\right)\right).
\end{equation}

Using that the observed flux density scales with the photon excess fraction, one can now argue that the observed surface brightness estimated by relation (\ref{eq:surfacebrightnessDefinition}) also scales with $I\left(O\right) = I_{\text{GR}}\left(O\right) \left(1+\mu_{\text{vio}}\right)$, where $I_{\text{GR}}\left(O\right)=I(S) \left(1+z\right)^{-4}$ is the classically observed surface brightness which one would receive from the intrinsic source surface brightness $I(S)$ in general relativity as discussed by \cite[Chap. 3]{Schneider1992}. For $\mu_{\text{vio}}=0$ this is the redshift corrected surface brightness conservation law, while $\mu_{\text{vio}}\neq0$ would clearly lead to unexpected variations in the surface brightness. Thus the relative surface brightness variation in area-metric lensing with respect to the surface brightness one would expect from classical general relativity is given by
\begin{equation}\label{eq:SurfacebrightnessFluctuationE}
\frac{\delta I}{I} = \frac{I (O) - I (O)_{\text{GR}} }{I (O)_{\text{GR}}} =  \mu_{\text{vio}}=\frac{3 \delta G M}{c^2}\left(\frac{\exp\left(-\eta \vert \bmath{r}_{{S}} -\bmath{r}_{{S}} \vert\right)}{\vert \bmath{r}_{{S}} -\bmath{r}_{{S}} \vert}-\frac{\exp\left(-\eta \vert \bmath{r}_{{O}} -\bmath{r}_{{M}} \vert\right)}{\vert \bmath{r}_{{O}} -\bmath{r}_{{M}} \vert}\right),
\end{equation}
where the lens is given by a point mass.

In the weak field limit considered here, which allows for linearity, it is possible to express this result (\ref{eq:SurfacebrightnessFluctuationE}) also for a continuous mass distribution $\rho_{\text{LSS}}(\bmath{r'})$ of the large scale structure (LSS), similar to \citet{rieser2020thesis} as 
\begin{equation}
\label{eq:singlesourceYukawa}
\frac{\delta I}{I}\left(\bmath{r}_{{S}},\bmath{r}_{{O}} \right)= \frac{3 \delta G }{c^2}\int \mathrm{d}^3 r'\:\rho_{\text{LSS}}(\bmath{r}')\left( \frac{1}{\vert \bmath{r}_{{S}} -\bmath{r}' \vert} \exp(-\eta \vert \bmath{r}_{{S}} - \bmath{r}'\vert) - \frac{1}{\vert \bmath{r}_{{O}} - \bmath{r}' \vert} \exp(-\eta \vert \bmath{r}_{{O}} - \bmath{r}'\vert) \right) =-3 \delta \left(\frac{\Phi_{{Y}}(\bmath{r_S})}{c^2} -\frac{\Phi_{{Y}}(\bmath{r_O})}{c^2}\right),
\end{equation}
with $K(\bmath{r} - \bmath{r}') =  -G \exp(-\eta \vert \bmath{r} - \bmath{r}'\vert)/\vert \bmath{r} - \bmath{r}' \vert$ as kernel function for the Yukawa potential $\Phi_{{Y}}\left(\bmath{r}\right)$. Now, we make the following assumption, namely that if we average over different sources and directions, and consider the observer as an ideal FLRW-observer comoving with the Hubble flow, then this observer would not measure the Yukawa correction potential $\Phi_{{Y}}(\bmath{r_O})$. Then, assuming standard $\Lambda$CDM cosmology the line of sight projected surface brightness fluctuation averaged over the source distribution of a flux limited survey according to relation (\ref{eq:eucliddistribution}) is given by
\begin{equation}\label{eq:ansatzAverage}
\overline{\frac{\delta I}{I}}\left(\bmath{\theta}\right)= - 3 \delta \int_0^{\chi_H} \mathrm{d} \chi \: p\left(z(\chi)\right)  \Theta\left(\chi_{A} - \chi\right)  \frac{{H}\left(\chi\right)}{c} D_+(a)  \frac{\Phi_{{Y}}(\chi\bmath{\theta}, \chi)}{c^2}.
\end{equation}
Here, $\chi_A$ is the source position up to which the line of sight integral is performed, thus dividing the line of sight into various extended bins, while $D_+(a)$ scales the potential as linear growth factor. The Euclidean distances of expression (\ref{eq:singlesourceYukawa}) were converted to comoving distances $\chi$ due to conformal flatness. One hereby makes the standard assumption that on cosmological scales the background geometry on which the perturbations, i.e. the so-called clump contributions \citep{Schneider1992}, are set is not Minkowskian, but an FLRW background, which is however conformally flat. Indeed, \citet{Duell2020thesis} and \citet{Fischer2017Thesis} could show that even in pre-metric cosmology the resulting area-metric is induced by a FLRW metric. Even though the full result, which these authors obtained, also contains a possible extra scale factor, whose dynamics is imprinted in a set of refined Friedmann equations \citep[see][for details]{Duell2020thesis}, we stick to $\Lambda$CDM cosmology evolving with only one scale factor $a(t)$ with the conventional Friedmann equations \citep[which do exist in area-metric spacetimes, see][]{duell2020symmetric} for simplicity and focus on the effect of point mass perturbations in the weak field limit of an area-metric geometry on lensing as a starting point. 

Next, we want to find the statistical spectra for these surface brightness fluctuations in area-metric lensing, and correlate them to the surface brightness fluctuations caused by intrinsic alignment. Therefore, we need to express the surface brightness fluctuations in Fourier space using the according Poisson equation. Due to the linearity one can find a set of coupled comoving Poisson equations for the effective potential $\Phi_{\text{eff}}$ (\ref{eq:effPotential}) according to
\begin{equation} \label{eq:Poissoneff}
\Delta_{\chi} a^{-2} \frac{\Phi_{\text{eff}}\left(\chi \bmath{\theta}, \chi\right)}{c^2} - \eta^2 \left(\frac{\Phi_{\text{eff}}\left(\chi \bmath{\theta}, \chi\right)}{c^2} - \frac{\Phi_{{N}}\left(\chi \bmath{\theta}, \chi\right)}{c^2} \right) = \left(1+\delta \right)\frac{3 \Omega_{m_0}}{2 \chi_H^2} \delta_c\left(\chi \boldsymbol{\theta}, \chi \right) a^{-3}, \quad \Delta_{\chi} a^{-2} \frac{\Phi_{{N}}\left(\chi \bmath{\theta}, \chi\right)}{c^2} =  \frac{3 \Omega_{m_0}}{2 \chi_H^2} \delta_c\left(\chi \boldsymbol{\theta}, \chi \right) a^{-3},
\end{equation}
similar to \citet{rieser2020thesis}. Here, $\Phi_{{N}}$ is the Newtonian potential, and $\Delta_\chi$ the spatial Laplacian in comoving coordinates. In Fourier space the according Fourier transform $\tilde{\Phi}_{\text{eff}}\left( k, \chi\right)$ of the effective potential is thus given 
\begin{equation}\label{eq:potentialFou}
\frac{\tilde{\Phi}_{\text{Def}}}{c^2}\left( k, \chi \right) = - \left(k^2+\eta^2 a^2\right)^{-1} \left(1+\delta+a^2\eta^2 k^{-2}  \right)\frac{3 \Omega_{m_0}}{2 \chi_H^2} \tilde{\delta}_c\left( k, \chi\right) a^{-1},
\end{equation}
with $k=\ell/\chi$ and $\Delta_{\chi} \rightarrow - k^2$. This result reproduces the Newtonian Poisson equation for $\eta \rightarrow 0$ and $\delta \rightarrow 0$ as expected. Also one can read off from (\ref{eq:potentialFou}) that the Yukawa corrections to the potential become negligible in the early universe when the scale factor squared is $a^2 \ll 1$, such that one can assume structure formation to be well described by the typical power spectrum $P_{\delta_c \delta_c}(k)\varpropto k^{n_S} T(k)^2$, with $T(k)$ again as the transfer function for either linear \citep{Bardeen1986} or non-linear structure formation \citep{Smith2003}, for density fluctuations to first order without Yukawa corrections. This will become important for evaluating the spectra. For a detailed discussion on how the Yukawa correction can further influence structure formation and thus the spectrum \citep[see][for details]{rieser2020thesis}. However, we can assume here that further modifications to the power spectrum would be of order $\mathcal{O}\left(\delta\right)$, while the surface brightness fluctuation itself is of order $\mathcal{O}\left(\delta\right)$, such that contributions of order $\mathcal{O}\left(\delta^2\right)$ are neglected here.

Similarly to~(\ref{eq:Poissoneff}) the comoving Poisson equation for the pure Yukawa potential $\Phi_Y$ is given by \begin{equation}
\left(\Delta_\chi a^{-2}- \eta^2  \right)\frac{\Phi_{{Y}}(\chi\bmath{\theta}, \chi)}{c^2} = \frac{3 \Omega_{m_0}}{2 \chi_H^2} \delta_c(\chi\bmath{\theta}, \chi) a^{-3},
\end{equation}
with  Fourier transform $\tilde{\Phi}_Y$:
\begin{equation}
\frac{\tilde{\Phi}_{{Y}}\left(k, \chi\right)}{c^2} = 
-\left( k^2+\eta^2 a^2  \right)^{-1}\frac{3 \Omega_{m_0}}{2 \chi_H^2} \tilde{\delta}_c\left( k, \chi\right) a^{-1}.
\end{equation}
Thus the surface brightness fluctuations in Fourier space are given by
\begin{equation}
\label{eq:muviolatefourier}
\widetilde{\overline{\frac{\delta I}{I}}}_{{Y}}(\ell) = 
3 \delta \int_0^{\chi_H} \mathrm{d} \chi\: W_{Y,A}\left(\chi\right) \left( k^2+\eta^2 a^2  \right)^{-1}\frac{3 \Omega_{m_0}}{2 \chi_H^2} \tilde{\delta}_c(k, \chi),
\end{equation}
with according weighting function
\begin{equation}
W_{{Y},A} \left(\chi\right) =  p\left(z(\chi)\right)  \Theta\left(\chi_{A} - \chi\right)  \frac{{H}\left(\chi\right)}{c} \frac{D_+(a)}{a}. 
\end{equation}

Finally the auto-correlation $\left\langle \widetilde{\overline{\delta I}}/I_{{Y}}(\ell) \: \widetilde{\overline{\delta I}}/I_{{Y}}(\ell') \right\rangle = (2 \pi)^2 \delta_D \left(\ell - \ell'\right) C_{A B}^{\delta \bar{I} /\bar{I}_Y  \, \delta \bar{I} /\bar{I}_Y}(\ell) $ of intrinsic surface brightness fluctuations due to the modified Etherington distance duality relation, which scales with Yukawa refinements in the gravitational potential, is defined by the $YY$-spectrum $C_{A B}^{\delta \bar{I} /\bar{I}_Y  \, \delta \bar{I} /\bar{I}_Y}(\ell) $ which is given by
\begin{equation}\label{eq:YY_Etherington}
YY: \,\, C_{A B}^{\delta \bar{I} /\bar{I}_Y  \, \delta \bar{I} /\bar{I}_Y}(\ell) = 9 \delta^2 \int_0^{\chi_H} \frac{\mathrm{d} \chi}{\chi^2}\: W_{{Y},A} \left(\chi\right) W_{{Y},B} \left(\chi\right) \frac{9 \Omega_{m_0}^2}{4 \chi_H^4} \left( k^2 +\eta^2 a^2  \right)^{-2} P_{\delta_c \delta_c}(k).
\end{equation}

Here, the Limber approximation \citep{limber} as well as the homogeneity and isotropy of the Gaussian distributed density fluctuations was again used, and $\delta_D \left(\ell - \ell'\right)$ denotes the delta distribution. Finally, one can also similarly find the $YI$ cross-correlation spectrum between the intrinsic surface brightness fluctuations derived in the last section \ref{sect_fluctuations_intrinsic} and the surface brightness fluctuations due to the photon excess in an area-metric spacetime. It is given by
\begin{equation}\label{eq:GI_Etherington}
GI: \,\, C_{A B}^{\delta \bar{I} /\bar{I}_Y  \, \delta \bar{I} /\bar{I}}(\ell) = - 3 \ell^2 \delta D_{IA} S_{\text{Sérsic}}(n) \int_0^{\chi_H} \frac{\mathrm{d} \chi}{\chi^2}\: D_{IA} S_{\text{Sérsic}}(n)  W_{\varphi,A} (\chi) W_{Y,B} (\chi) \frac{9 \Omega_{m_0}^2}{4 \chi_H^4} k^{-2} \left(k^2 +\eta^2 a^2  \right)^{-1} P_{\delta_c \delta_c}(k),
\end{equation}
where the negative sign stems from the convention that lensing effects are usually anti-correlated to intrinsic alignment effects as discussed for instance in \citet{ghosh2020intrinsic}. Whether or not there is an anti-correlation between the intrinsic surface brightness fluctuation and the surface brightness fluctuation due to area-metric corrections is however unclear and especially depends on the sign of $\delta$. All Limber-type integrations implicitly assume the relation $\ell = k\chi$ between multipole and wavenumber.

\section{Numerical results on the surface brightness fluctuation spectra}
\label{sect_numerical}
After having derived the $II$, $YI$ and $YY$ spectra 
(\ref{eq:IIspectrumE}), (\ref{eq:YY_Etherington}) and (\ref{eq:GI_Etherington}) we need to estimate their shape and amplitude depending on different choices for the area-metric constants $\eta$ and $\delta$. While in recent works, for instance in \citet{Henrichs21, rieser2020thesis} these constants were determined via Yukawa refinements in the gravitational potential of the Milky Way, we will estimate them in terms of inverse multiples of the Hubble length, for otherwise they might have already been detected as significant via lensing. Also we want to estimate which values the coupling $\delta$ and the inverse range $\eta$ need to have to be bounded from above by the surface brightness fluctuation spectrum due to intrinsic alignment. Thus we make the following dimensional argument via the comoving Poisson equation for the Yukawa potential in Fourier space given by
$\delta \tilde{\Phi}_{{Y}}\left(k, \chi\right)/c^2= -\left( k^2+\eta^2 a^2  \right)^{-1}\, 3 \Omega_{m_0}/2 \, \, \delta/ \chi_H^2\, \tilde{\delta}_c\left( k, \chi\right) a^{-1}$. Firstly, one can see that the parameter $\eta$ has units of inverse length as the wavenumber $k$, such that we choose multiples of the Hubble length $\chi_H$ where new physics effects like area-metric corrections start to become important, thus we set $\eta^{-1} = m\chi_H$. Secondly, the parameter $\delta$ as coupling of the Yukawa correction $\tilde{\Phi}_{{Y}}$ to the Newtonian potential scales with inverse Hubble length squared $\chi_H^{-2}$. Thus, for Yukawa corrections becoming important at a scale of $m \chi_H$ then $\delta$ should scale with $1/m^2$.

\begin{figure}
	\centering
	\includegraphics[scale=0.45]{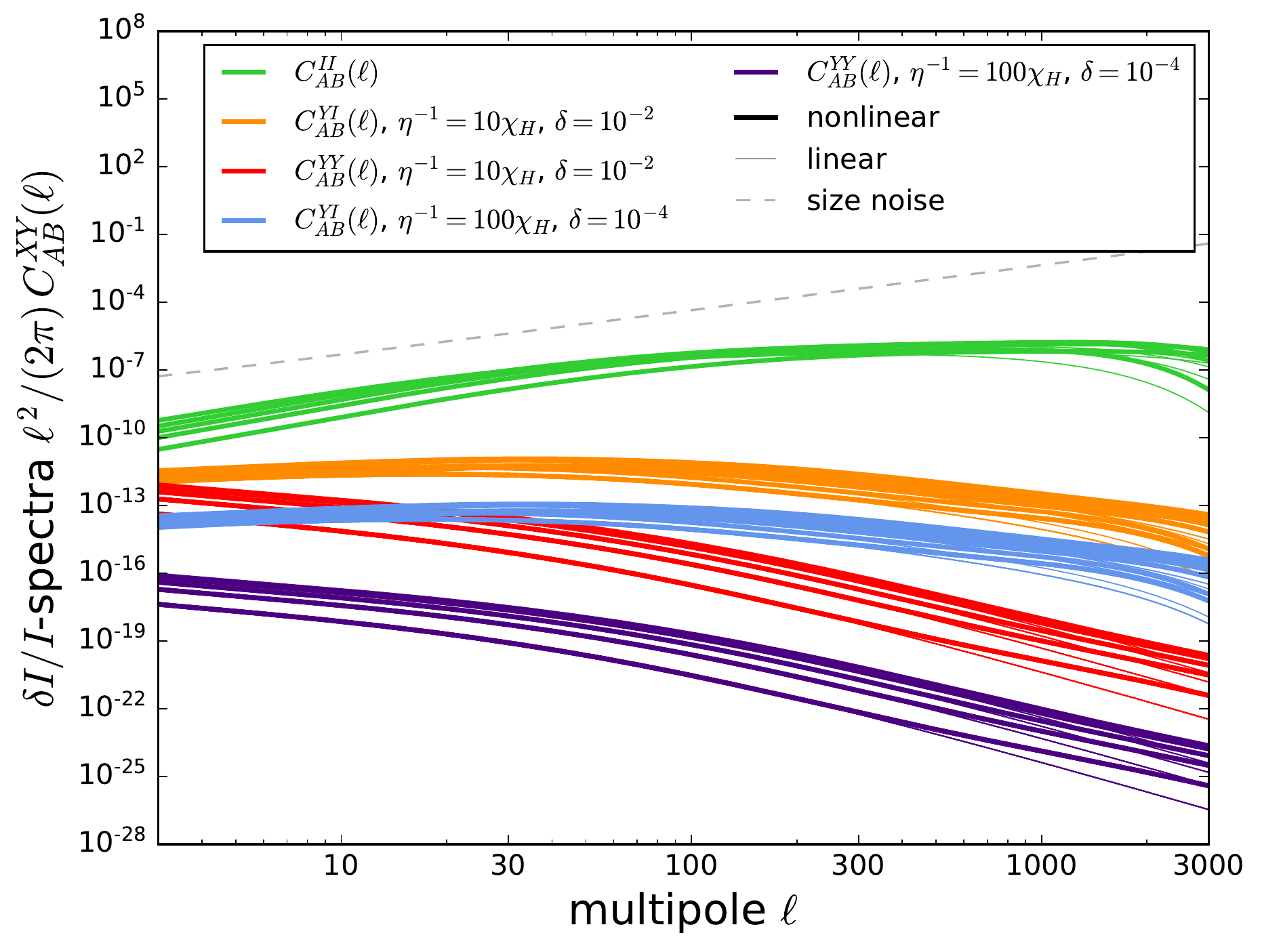}
	\caption{\textbf{Surface brightness variation spectra for Sérsic index $\boldsymbol{n=1}$ for a $\boldsymbol{5}$-bin tomography:} The amplitudes of the $YI$ spectra (denoted as $C_{AB}^{YI}(\ell)$ ) and $YY$ spectra (denoted as $C_{AB}^{YY}(\ell)$) are compared to the $II$ spectrum (depicted in green and denoted as $C_{AB}^{II}(\ell)$) for different choices of Yukawa refinement parameters with $\eta^{-1} = 10 \chi_H$ with $\delta=10^{-2}$ (orange curve for cross-correlation, red curve for auto-correlation), and $\eta^{-1} = 100 \chi_H$ with $\delta=10^{-4}$ (blue curve for cross-correlation, purple curve for auto-correlation). }
	\label{fig:CompareSpectran=1}
\end{figure}

In figure \ref{fig:CompareSpectran=1} the surface brightness fluctuation spectra are shown for a Sérsic index of $n=1$ corresponding to the exponential profile, and different choices for $\eta$ and $\delta$ with $\eta^{-1}=10 \chi_H$ with $\delta=10^{-2}$, and $\eta^{-1}=100 \chi_H$ with $\delta =10^{-4}$ respectively. We choose the exponential profile, corresponding for instance to dwarf galaxies, for simplicity, while noting that one could for instance also use higher Sérsic indices up to $n=4$ for a de Vaucouleurs profile for typical elliptical galaxies. This would not change the shape of the $II$- and $YI$-spectra qualitatively, but only their amplitude would be down scaled by the Sérsic-index dependent pre-factor $S_{\text{Sérsic}}(n)$ within the alignment parameter: While for an exponential profile this would be of magnitude $S_{\text{Sérsic}}(1)\approx 1$, we find $S_{\text{Sérsic}}(4)\approx 10^{-2}$ for a de Vaucouleurs-profile. While the thick lines in figure \ref{fig:CompareSpectran=1} represent spectra based on a model for non-linear structure formation according to \citet{Smith2003}, we assumed a linear power spectrum for the thin lines. In both cases a Gaussian smoothing is introduced for the intrinsic alignment spectrum and cross correlation spectrum on galaxy scales, cutting off tidal field fluctuations below the spatial scale of galaxies.

For the parameters chosen the $II$-spectrum is expected to have the largest signal compared to the $YI$ spectrum and the $YY$ spectrum, which has the lowest amplitude. Furthermore, the $II$ spectrum increases for smaller structures, while the $YI$ and the $YY$ spectra decrease, indicating that surface brightness fluctuations due to the modification of the Etherington distance duality on an area-metric background is rather a large scale effect and suppressed on small scales. Also, the shape of the $YI$ spectra and the $YY$ spectra do not vary significantly with different $\eta$, while their amplitudes are more sensitive to changes in the parameter $\delta$. The noise is given in terms of the size error since, as we discussed previously the surface brightness fluctuations $\delta I/I$ are proportional to the size fluctuations $\delta s/s$, typically modelled as 
\begin{equation}
\left(N_{\text{noise}}\right)_{A B} = \sigma^2_{\text{size}} \frac{n_{\text{tomo}}}{\bar{n}} \delta_{A B},
\end{equation}
with $n_{\text{tomo}}$ as the number of bins, $\sigma_{\text{size}}=0.8$ the size noise for elliptical galaxies and $\bar{n} = 3.545\times 10^8 \text{sr}^{-1}$ as the observable number density for surveys like Euclid \citep{laureijs2011euclid}.

\begin{figure}
	\centering
	\includegraphics[scale=0.45]{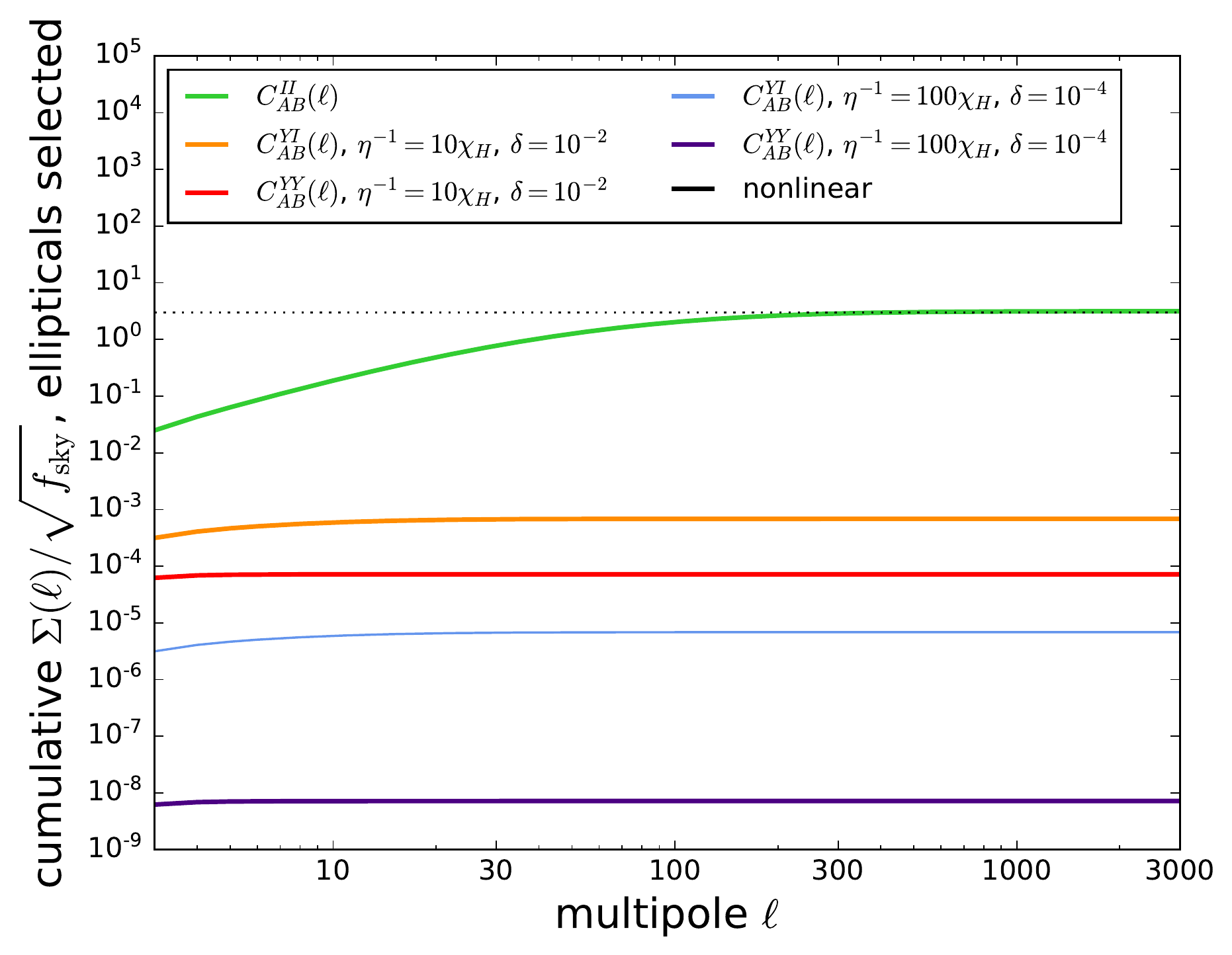}
	\caption{\textbf{Cumulated signal to noise ratio $\boldsymbol{\Sigma}$ for the surface brightness variation spectra in a $\boldsymbol{5}$-bin tomography, with Sérsic index $\boldsymbol{n=1}$, and elliptical galaxy sample:} The cumulated signal to noise ratio of the $C_{AB}^{II}(\ell)$ spectrum (depicted in green) is compared to the cumulated signal of noise ratios of the $C_{AB}^{YI}(\ell)$ spectra and $C_{AB}^{YY}(\ell)$ spectra for different choices of parameters $\eta$ and $\delta$: The orange curve (cross-correlation) and the red curve (auto-correlation) depict $\Sigma(\ell)$ for $\eta^{-1} = 10 \chi_H$ and $\delta=10^{-2}$. The blue curve (cross-correlation) and the purple curve (auto-correlation) depict $\Sigma(\ell)$ for and $\eta^{-1} = 100 \chi_H$ with $\delta=10^{-4}$. The factor $\sqrt{f_{\text{sky}}}$ corrects for incomplete sky coverage.}
	\label{fig:S2Nell1E}
\end{figure}

\begin{figure}
	\centering
	\includegraphics[scale=0.45]{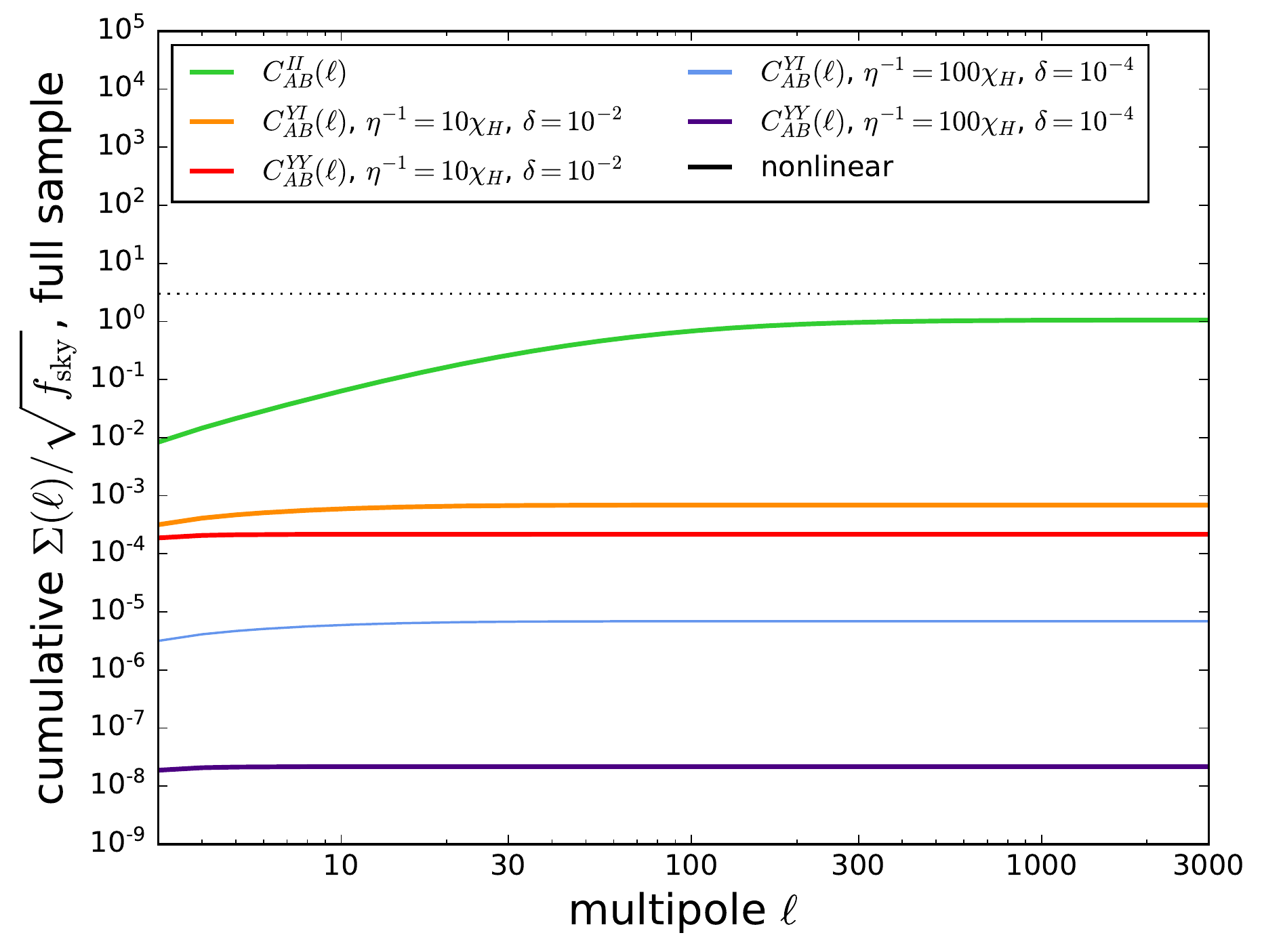}
	\caption{\textbf{Cumulated signal to noise ratio $\boldsymbol{\Sigma}$ for the surface brightness variation spectra in a $\boldsymbol{5}$-bin tomography, with Sérsic index $\boldsymbol{n=1}$, and full galaxy sample:} The cumulated signal to noise ratio of the $C_{AB}^{II}(\ell)$ spectrum (depicted in green) is compared to the cumulated signal of noise ratios of the $C_{AB}^{YI}(\ell)$ spectra and $C_{AB}^{YY}(\ell)$ spectra for different choices of parameters $\eta$ and $\delta$: The orange curve (cross-correlation) and the red curve (auto-correlation) depict $\Sigma(\ell)$ for $\eta^{-1} = 10 \chi_H$ and $\delta=10^{-2}$. The blue curve (cross-correlation) and the purple curve (auto-correlation) depict $\Sigma(\ell)$ for and $\eta^{-1} = 100 \chi_H$ with $\delta=10^{-4}$. The factor $\sqrt{f_{\text{sky}}}$ corrects for incomplete sky coverage.}
	\label{fig:S2Nfull1E}
\end{figure}

To estimate the measurability of these spectra with a Euclid-like survey we depict their according cumulated signal to noise ratios in figures \ref{fig:S2Nell1E} and \ref{fig:S2Nfull1E}. The cumulated signal to noise ratio $\Sigma$ is hereby given by 
\begin{equation}
\Sigma^2 = \sum_\ell \frac{2\ell+1}{2} \text{tr}\left(\mathcal{C}^{-1} S \, \mathcal{C}^{-1} S   \right)\,,
\end{equation}
\citep{tegmark_1997,hu_weak_1999} with the signal $S(\ell)$ as either the $II$ spectrum $ C_{A B}^{\delta \bar{I} /\bar{I}  \, \delta \bar{I} /\bar{I}}(\ell)$, $YI$ spectum $ C_{A B}^{\delta \bar{I} /\bar{I}_Y  \, \delta \bar{I} /\bar{I}}(\ell)$ or $YY$ spectrum $ C_{A B}^{\delta \bar{I} /\bar{I}_Y  \, \delta \bar{I} /\bar{I}_Y}(\ell)$ and the covariance matrix $\mathcal{C}$ can be expressed in terms of the bin components
\begin{equation}
\mathcal{C}_{A B}(\ell) = 
 C_{A B}^{\delta \bar{I} /\bar{I}  \, \delta \bar{I} /\bar{I}}(\ell) +  C_{A B}^{\delta \bar{I} /\bar{I}_Y  \, \delta \bar{I} /\bar{I}}(\ell) +  C_{A B}^{\delta \bar{I} /\bar{I}_Y  \, \delta \bar{I} /\bar{I}_Y}(\ell) + \left(N_{\text{noise}}\right)_{A B}.
\end{equation}

While in figure \ref{fig:S2Nell1E} the signal strength is depicted for a sample of elliptical galaxies, the full galaxy sample also including spirals is considered in 	\ref{fig:S2Nfull1E}. Similar to the discussion in \citet{ghosh2020intrinsic,Giesel_2021} the signal strength for the intrinsic alignment part of the spectra is suppressed by a factor of $q=1/3$ for the $YI$ spectrum and $q^2$ for the $II$ spectrum, since we assumed the alignment model to apply to ellipticals only, working with a rough number of one third for the fraction of ellipticals in the entire galaxy sample. However, due to the larger sample size the total Poissonian error is also $1/3$ of the noise expected for elliptical galaxies. One can observe that observability is only possible for the $II$-signal for elliptical galaxies, where $\Sigma(\ell)$ exceeds he cumulated signal to noise ratio of $\Sigma(\ell)\geq 3$ for multipole orders above $l \geq 100$. The cumulated signal to noise ratios for the $YI$ and $YY$ signal are strongly suppressed however, such that with the chosen parameters $\eta$ and $\delta$ no detection would be possible.

\begin{figure}
	\centering
	\includegraphics[scale=0.45]{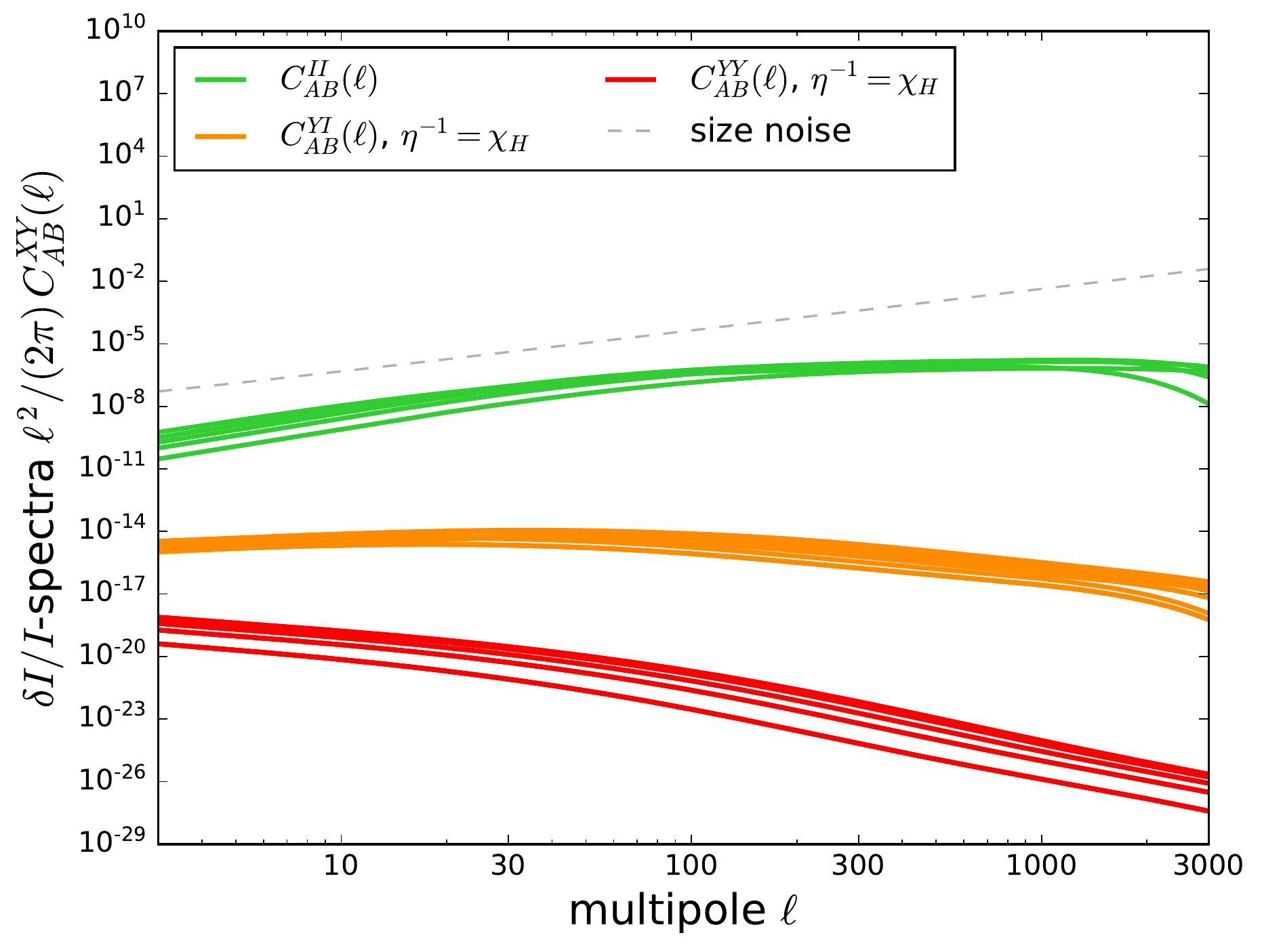}
    \caption{\textbf{Surface brightness variation spectra for Sérsic index $\boldsymbol{n=1}$ and a $\boldsymbol{5}$-bin tomography and fixed $\boldsymbol{\delta = 10^{-5}}$:} 
    \\
    The amplitudes of the $YI$ spectrum (denoted as $C^{YI}_{AB} (\ell)$ and depicted in orange) and the $YY$ spectrum (denoted as $C^{YY}_{AB} (\ell)$ and depicted in red) are compared to the $II$ spectrum $C^{II}_{AB} (\ell)$ shown in green for $\eta^{-1} = \chi_H$. The spectra are based on a non-linear power spectrum only. }
	\label{fig:SpectraFixedDelta}
\end{figure}

\begin{figure}
    \centering
    \includegraphics[scale=0.25]{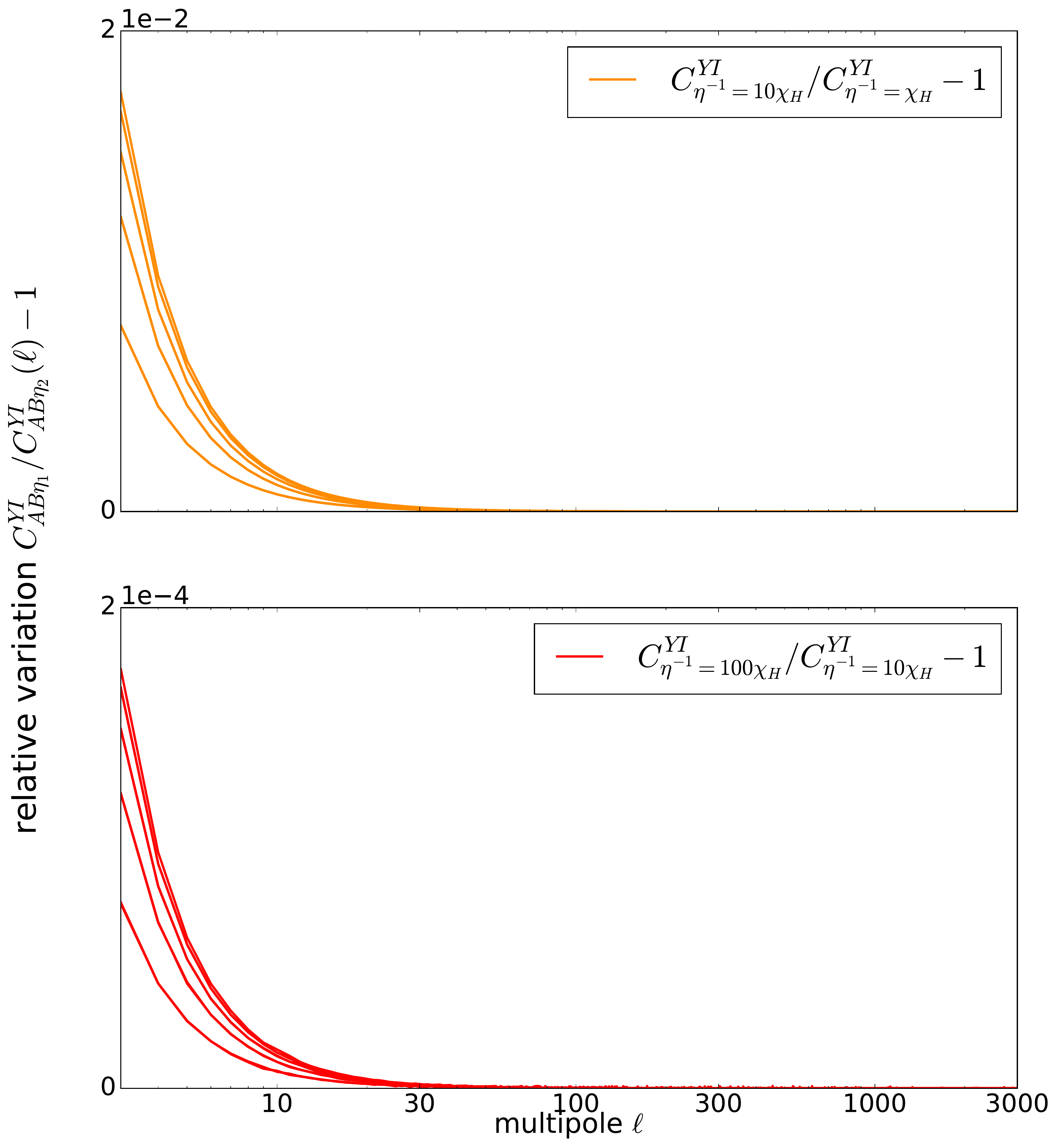} 
    \caption{\textbf{Ratio of the $\boldsymbol{C^{YI}_{AB}(\ell)}$ spectra for different values of $\boldsymbol{\eta_1}$ and $\boldsymbol{\eta_2}$ with $\boldsymbol{\delta=10^{-5}}$:} 
    \\In the first plot the ratio $C^{YI}_{AB;\eta_1}\left(\ell\right)/C^{YI}_{AB;\eta_2}\left(\ell\right) -1$ is depicted in orange for a $5$-bin tomography with $\eta_1^{-1}=10\chi_H$ and  $\eta_2^{-1}=1\chi_H$. In the second plot the same is shown for values $\eta_1^{-1}=100\chi_H$ and $\eta_2^{-1}=10\chi_H$ in red. In both cases the ratio of the spectra is close to one and the difference decreases further for increasing multipole order $\ell$.}
	\label{fig:SpectraRelativeYIsub}
\end{figure}

\begin{figure}
    \centering
    \includegraphics[scale=0.25]{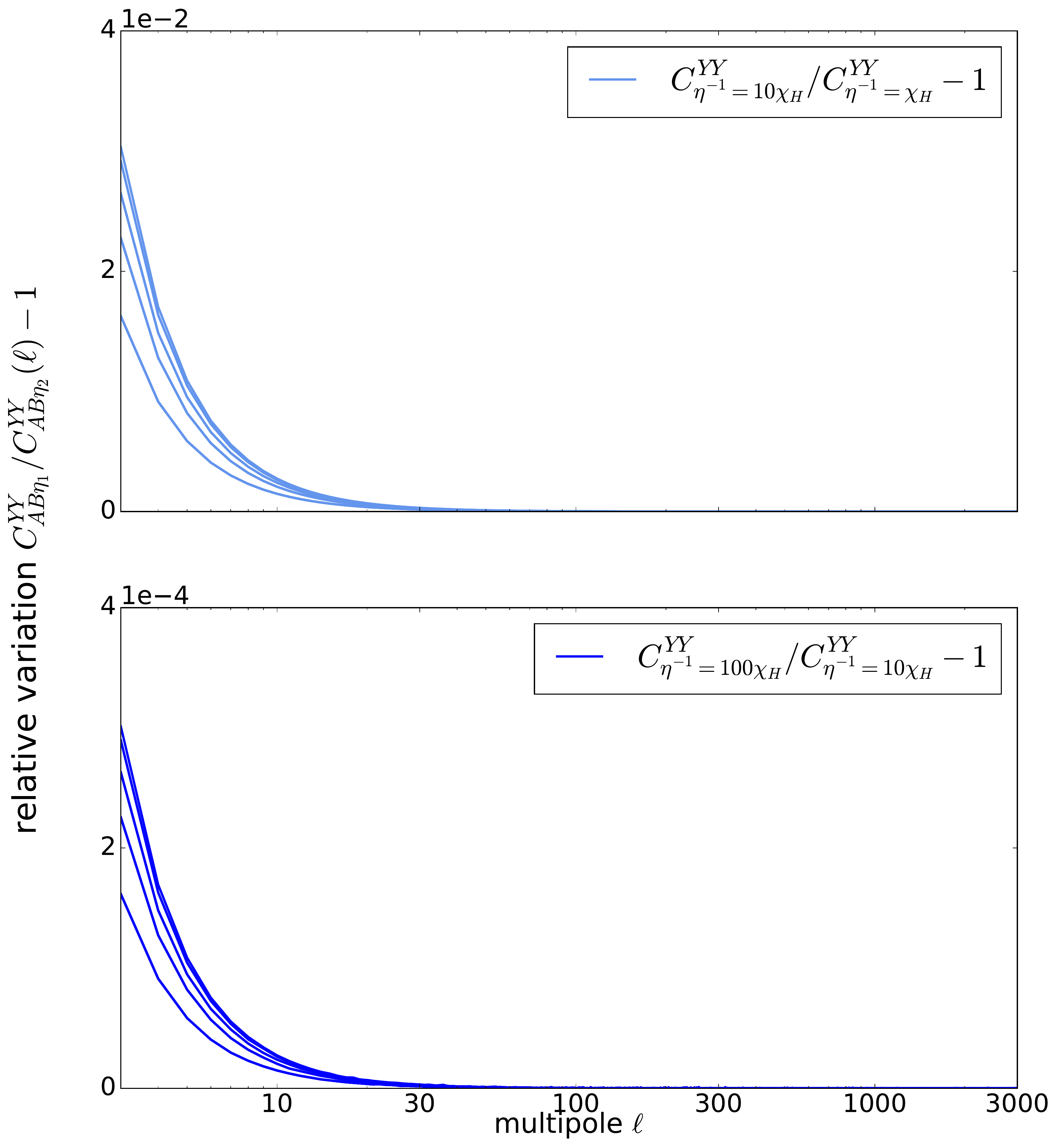}
    \caption{\textbf{Ratio of the $\boldsymbol{C^{YY}_{AB}(\ell)}$ spectra for different values of $\boldsymbol{\eta_1}$ and $\boldsymbol{\eta_2}$ with $\boldsymbol{\delta=10^{-5}}$:} 
    \\In the first plot the ratio $C^{YY}_{AB;\eta_1}\left(\ell\right)/C^{YY}_{AB;\eta_2}\left(\ell\right) -1$ is depicted in light blue for a $5$-bin tomography with $\eta_1^{-1}=10\chi_H$ and  $\eta_2^{-1}=1\chi_H$. In the second plot the same is shown for values $\eta_1^{-1}=100\chi_H$ and $\eta_2^{-1}=10\chi_H$ in dark blue. In both cases the ratio of the spectra is close to one and the difference decreases further for increasing multipole order $\ell$ similar to figure \ref{fig:SpectraRelativeYIsub}.}
    \label{fig:SpectraRelativeYYsub}
\end{figure}

Furthermore, we want to study how different choices for the range parameter $\eta$ would affect the shape of the spectra for fixed coupling parameter $\delta$. We hereby set $\delta = 10^{-5}$ corresponding to the magnitude of the relative uncertainty of the gravitational constant $G$ according to CODATA \citep{CODATA} with $\Delta G/G =2.2 \times 10^{-5}$. This is a reasonable choice since for vanishing $\eta \rightarrow 0$ the exponential in the Yukawa correction of the potential would go to one, such that the correction itself would just become an off-set to the Newtonian potential as also remarked in \citet{rieser2020thesis}. The according spectra are depicted in figure \ref{fig:SpectraFixedDelta} with $\eta =1/\chi_H$ as inverse length scale. One can see that, while the amplitude of the $YI$ and the $YY$ spectra is decreased compared to figure \ref{fig:CompareSpectran=1} due to the smaller value of $\delta$, the shape of the spectra does not change significantly with a different value of $\eta$. 
This can be specified in more detail by considering the relative $YI$ and $YY$ spectra for different values of $\eta$ and fixed $\delta=10^{-5}$ as shown in figures \ref{fig:SpectraRelativeYIsub} and \ref{fig:SpectraRelativeYYsub}. Here the relative amplitudes $C^{YI}_{AB;\eta_1}/C^{YI}_{AB;\eta_2}-1$ are depicted for different choices of $\eta$, respectively $C^{YY}_{AB; \eta_1}/C^{YY}_{AB;\eta_2}-1$. As the spectra depend on $\eta$ in a non-linear way with $C^{YI}_{AB}\left(\ell\right) \varpropto \left(1+\eta^2 a^2\right)^{-1}$ from (\ref{eq:GI_Etherington}) and $C^{YY}_{AB}\left(\ell\right) \varpropto \left(1+\eta^2 a^2\right)^{-2}$ from (\ref{eq:YY_Etherington}) one can make the following approximation for their ratios since $\eta$ is small:
We estimate that
\begin{equation}
C^{YI}_{AB}\left(\eta_1\right) \varpropto \left(1+\eta_1^2 a^2\right)^{-1} \approx 1- \eta_1^2,
\quad C^{YI}_{AB}\left(\eta_2\right) \varpropto \left(1+\eta_2^2 a^2\right)^{-1} \approx 1- \eta_2^2,
\quad \quad \Rightarrow \,\, \frac{C^{YI}_{AB}\left(\eta_1\right)}{C^{YI}_{AB}\left(\eta_2\right)}-1 \approx \frac{1-\eta_1^2}{1-\eta_2^2} -1 \approx \eta_2^2 -\eta_1^2 = \Delta \eta^2,
\end{equation}
and analogously
$C^{YY}_{AB}\left(\eta_1\right)/C^{YY}_{AB}\left(\eta_2\right)-1 \approx  \Delta \eta^2,
$
such that the plotted ratios directly measure the quadratic difference in $\eta$.
One can see that in both cases the relative amplitudes are very close to one, and thus the curves depicted in figures \ref{fig:SpectraRelativeYIsub} and \ref{fig:SpectraRelativeYYsub} start off at magnitudes of $10^{-2}$
($YI$) to $10^{-4}$ ($YY$) at low multipole orders and decrease even further for increasing multipole orders. Thus, the effect on the spectra for different values of $\eta$ is very small, and not significant on small scales.

\section{summary}
\label{sect_summary}
In this work, we have performed a study of surface brightness fluctuations, arising on one hand due to the violation of the Etherington distance duality relation in gravitational lensing in area-metric spacetimes, and on the other hand due to direct tidal interaction of the source with the environment, i.e. through an intrinsic alignment effect. The main goal was to quantify whether the Etherington duality violating effect caused by an area-metric spacetime structure can be distinguished from the classical physics analog caused by intrinsic alignment. It turned out that the signal of surface brightness fluctuations due to intrinsic alignment dominates the signal of the analog effect in non-metric gravitational lensing by several orders of magnitude. Thus, while surface brightness fluctuations due to intrinsic alignment effects were shown to be measurable with surveys like Euclid, the area-metric Etherington distance duality effect is not resolvable in comparison.

In the first case, we find that the intrinsic brightness fluctuation has the same proportionality to $\Delta\Phi/c^2$ as that of the intrinsic size fluctuation, which enables us to compute the auto-correlation spectra ($II$-spectra) for the brightness fluctuation in terms of the size fluctuation, which we have already obtained previously in \citet{ghosh2020intrinsic}. While calculating the surface brightness of the galaxy, we have considered the total number of stars and therefore the light output to be constant. However, if we take additional star formation due to tidal interaction into account \citep{renaud2010dynamics}, we will encounter a change in the flux which can lead to an intrinsic magnification effect, which is however beyond the scope of this work. 

Furthermore, in order to study the surface brightness violations due to a modified Etherington distance duality relation in a weakly birefringent spacetime, we first see that the usual Newtonian potential acquires a Yukawa correction term, in which the parameters $\eta$ and $\delta$ correspond to the inverse range of the Yukawa interaction and the strength of the coupling to the otherwise Newtonian potential, respectively. In area-metric geometries, the photon number is still covariantly conserved, however, with respect to a different volume measure compared to metric geometry, such that effectively the photons are focused differently compared to standard gravitational lensing. This leads to an excess fraction of photons which can be obtained in terms of the parameters $\eta$ and $\delta$. The surface brightness fluctuation can then be easily obtained for point mass sources, which can be effectively extended for continuous mass distributions. We then compute the auto-correlation spectra ($YY$-spectra) and cross-correlation spectra ($YI$-spectra) with the intrinsic alignment effect for these fluctuations for different parameterisations of $\eta$ and $\delta$, and estimate the shape and the amplitude of the same for different choices of the said parameters. Importantly, these parameters are a priori not constrained by the underlying theory of constructive gravity, but since the parameter $\eta$ is seen to have units of inverse length, a natural choice is to express it in terms of multiples of the Hubble length as a natural scale for cosmology, while $\delta$ scales with inverse Hubble length squared. The values are thus chosen such that the effect of the area-metric refinements becomes important on large scales, and also such that they are bounded from above by the intrinsic brightness fluctuation spectrum. 

With the chosen values we find that the $II$-spectrum has the highest amplitude, and grows on smaller scales, in comparison to the $YI$ and $YY$ spectra which are low in amplitude to begin with and decrease even further on smaller scales. It is also seen that the shape of the $YI$ and $YY$ spectra are more sensitive to $\delta$ than to $\eta$. While in our work we study the effect of Yukawa corrections on cosmological scales to light propagtion, one should remark that there are also efforts to investigate Yukawa screenings of gravity on local scales \citep[see][for example]{rieser2020thesis, Henrichs21}. Also, while mechanisms to include Yukawa corrections into modified gravity appear in many examples of $f(R)$-gravity \citep[see][for reviews]{amendola_tsujikawa_2010,CLIFTON20121} this work especially focuses on the prediction of Etherington distance duality violation according to \citet{schuller2017etheringtons} which particularly occurs in area-metric gravity.

For computing the signal-to-noise ratio we consider two cases similar to \citet{ghosh2020intrinsic}, namely, the elliptical galaxies sample and the full galaxy sample. Our results are found to be optimistic as far as observations are concerned only in the case of elliptical galaxies for $II$-signals, that too for $\ell\geq 100$ where the cumulated signal-to-noise ratio $\Sigma(\ell)\geq 3$. For the $YI$ and $YY$ signal the cumulated signal is too low for any realistic observability. We also check the effect of different values of $\eta$ for a fixed value of $\delta=10^{-5}$ as uncertainty of the gravitational constant $G$ by computing the ratio of spectra, and find that the relative difference is prominent only on large scales and negligible on small scales.

In principle, for a generalised treatment of area-metric spacetimes, we need to keep in mind that there occur many fundamentally unknown parameters in the theory, for instance those arising due to a modification of the Friedmann equations. For a fully consistent study we would need to understand these different unknown quantities which is beyond the scope of the present work. Another important assumption we made was to derive the intrinsic surface brightness fluctuations within Newtonian gravity, recognizing that if the spacetime structure was indeed area-metric, Yukawa corrections would also be required for the intrinsic alignment model. However, with the parameter values chosen for $\delta$ and $\eta$ in this work the additions to the Newtonian potential would only become numerically relevant on cosmological scales important for weak lensing, but not for local scales relevant for intrinsic alignment. And even if these corrections were included, they would not be numerically resolvable within the uncertainty of the alignment parameter $D_{\text{IA}}$, which is only determined up to a factor of $10$. But for larger values of $\delta$ and $\eta$ which would allow to notice Yukawa corrections on galaxy scales, a refinement of the lensing potentials including Yukawa corrections should be studied as well. Hence, one can also go on in future work to construct a new alignment potential analogous to the lensing potential, which includes the violation to the Etherington distance duality relation, and eventually compute the include power spectrum according to \cite{rieser2020thesis} for the same.

Beyond that the next natural step would be to additionally investigate how relative galaxy orientations with respect to the line-of-sight can further affect the amplitude and shape of intrinsic surface brightness fluctuation spectra, similar to a study conducted by \cite{Krause_2010}: Indeed, intrinsic alignment affects whether galaxies are more likely to be observed face-on or edge-on, what will lead to an effective bias on the observed surface brightness, and it would be worthwhile to see how this affects the total $II$-spectrum discussed in this work. Furthermore, It would be interesting to study how this effect is compared to brightness fluctuations caused by an area-metric, Etherington duality violating, spacetime.

Furthermore, we did not take the Malmquist selection bias \citep{Malmquist25} into account in this work: For a flux limited survey the brightness of the sources seems to increase with higher redshift. This apparent paradox can be resolved by noting that only the brightest stars are able to pass the selection process. Indeed, the higher the redshift the fainter the total number of sources, and the brighter are those sources which can still be measured. How this bias might affect the surface brightness fluctuation spectra, both in intrinsic alignment and Etherington duality violating lensing, as a function of the redshift would be very interesting for future work. We assume that due to the Malmquist bias sources within underdense regions, which experience an effective decrease in surface brightness due to the intrinsic alignment model described here, might be selected less likely. This would lead to a selection bias towards sources in overdense regions. How the Malmquist bias would affect the shape of the surface brightness fluctuation spectra due to a violation of the Etherington duality depends on the sign of the parameter $\delta$, which is however not known so far. If it is negative equation (\ref{eq:ansatzAverage}) shows, that the relative brightness would increase, while for $\delta > 0$ it would decrease, and therefore sources would be selected less likely for higher redshifts.

Finally, we would like to comment on the contribution of magnification bias-density correlations in the spectra, which for the violation or photon excess term, is seen to be around seven orders of magnitude smaller than the total effective magnification bias contribution, and can hence be neglected.

As last point we remark that violations of the Etherington distance duality have also been studied in different contexts, for instance if gravity is coupled non-minimally to matter \citep{Azevedo:2021npm}, or if the photon number is not conserved due to exotic photon decays, or photon mixing with axion fields leading to absorption \citep{Bassett_2004}. Effectively, the functional form of the Yukawa-term in Eq.~(\ref{eqn_magic}) would partially catch such an effect, as the photon flux would decrease exponentially with distance, again typically on a length scale comparable to the Hubble-scale. As a final remark we point out that in our particular case intrinsic effects are far dominating over the modified gravitational lensing effect, which is rather unusual given that the opposite is true for virtually all lensing studies.

\section*{Acknowledgements}
ESG would like to thank Studienstiftung des Deutschen Volkes and ARI (Astronomisches Rechen-Institut) at ZAH (Center for Astronomy at the University of Heidelberg) for financial support, as well as the YRC-STRUCTURES for providing a travel fund to the University of Twente. BG thanks the Department of Science and Technology for DST-INSPIRE Faculty Fellowship. The authors would like to express their gratitude to Marcus Werner, Frederic P. Schuller, Florian Wolz, Alexander Wierzba, Maximilian Düll, and Hans-Martin Rieser for their interest, thorough discussion and critical remarks. The authors would like to thank the anonymous referee for their time to review this work, as well as their valuable remarks for improvement.

\section*{Data Availability}
There are no new data associated with this article.

\appendix

\bibliographystyle{mnras}
\bibliography{refs}

\bsp
\label{lastpage}
\end{document}